\documentstyle[11pt,paspconf,psfig]{article}

\begin{document}

\title{Radiospectra and Kinematics in Blazars}

\author{A.~P.~Lobanov, A.~Kraus, J.~A.~Zensus}
\affil{Max-Planck-Institut f\"ur Radioastronomie, Auf dem H\"ugel 69,
Bonn 53121, Germany}

\begin{abstract}

Broad band spectra of total and compact--scale radio emission from
blazars, used in combination with kinematic information inferred from
VLBI monitoring programs, can be applied successfully to better
constrain the models of compact jets. We discuss here a ``hands-on''
approach for tying together the kinematic and spectral properties of
radio emission from blazars.

\end{abstract}

% Keywords should be included, but they are not printed in the hardcopy.


% That's it for the front matter.  On to the main body of the paper.
% We'll only put in tutorial remarks at the beginning of each section
% so you can see entire sections together.

\section{Introduction}

Analytical models of parsec--scale jets, although being contested ever
stronger by numerical simulations, still provide a valuable means for
describing observational data, particularly when different aspects of
the jet physics (such as its spectral and kinematic properties) are
combined together in a single formulation, providing additional
constraints and checks for the model parameters. In this contribution,
we discuss how the well--established shock--in--jet model (``shock
model'' hereafter; see Marscher 1990, Marscher, Gear,
\& Travis 1991, for detailed discussions of the model)
can be reinforced by inclusion of kinematic information available from
VLBI observations.

\section {Model quantities}

In its most common formulation, the shock model predicts changes of
the turnover frequency, $\nu_{\rm m}$, and flux density, $S_{\rm m}$,
in the spectrum of radio emission associated with a shock. The jet
is usually assumed to have a constant opening angle, $\phi$, so that
the shock transverse dimension is proportional to the distance, $r$, at
which the shock is located. Other model parameters are expressed as
functions of $r$: the magnetic field $B\propto r^{-a}$, Doppler factor
$\delta\propto r^{b}$, and number density $N\propto r^{-n}$ (for a
power--law electron energy distribution, $N(\gamma)d\gamma \propto
\gamma^{-s}d\gamma$). The shock emission is dominated subsequently by
Compton, synchrotron and adiabatic losses. At each stage, the
predicted quantities are described by the following proportionalities:
$S_{\rm m}
\propto \nu_{\rm m}^\rho$ and $\nu_{\rm m} \propto r^{\varepsilon}$,
with $\rho = \rho(a,b,s)$ and $\varepsilon = \varepsilon(a,b,s)$ (for
complete evaluations of $\rho$ and $\varepsilon$, see Marscher 1990,
Lobanov \& Zensus 1999). Below, we describe how estimates of the power
index $b$ can be obtained from VLBI data, and give examples of
applying this approach to studying compact variable radio sources.

\section {Observable quantities}

Single dish observations yield light curves $S(t)$ at different
frequencies, which can be used to determine the evolution of spectral
turnover ($S_{\rm m},\nu_{\rm m}$), provided an adequate frequency
coverage and time sampling. VLBI monitoring programs allow to measure
relative proper motions, $\mu_{\rm app}(t)$, and (in exceptional
cases) also spectral changes of enhanced emission regions detected in
parsec--scale jets. From the measured $\mu(t)$, the jet apparent
speeds, $\beta_{\rm app}(t)$, and Doppler factors, $\delta(t)$, can be
reconstructed, with necessary assumptions made about the jet
kinematics. In the simplest case, the jet Lorentz factor, $\gamma$, 
can be taken constant, and $\delta(t)$ is then described by changes of the
jet viewing angle, $\theta(t)$. For more complicated cases, $\gamma(t)
= \gamma_{\rm min}(t) = [1+\beta_{\rm app}(t)]^{0.5}$ can be assumed,
or even complete kinematic settings can be postulated (e.g.  a helical
trajectory, as has been done, for instance, by Roland et al. 1994).

\section{Relations between the jet spectrum and kinematics}

Once the form of $\delta(t)$ has been determined, we can evaluate
$b(t)$. Since variations of $\delta(t)$ are not necessarily monotonic,
we resort to determining $b(t)$ locally, so that
\begin{equation}
b(t) = \frac{\log[\delta(t+dt)/\delta(t)]}{\log[r(t+dt)/r(t)]}
\end{equation}
We then select a timerange, ($t_1,t_2$), during which the changes of
$b(t)$ are small enough to approximate $b(t_1)\approx b(t_2)\approx b$.
Fitting the observed spectral turnover data, we can obtain the 
turnover points at the respective epochs, and evaluate the absolute
location of the shock at the epoch $t_1$:
\begin{equation}
r_1 = \left( \frac{1+z}{\delta_1 c \Delta t} \int_{1}^{r_u} 
\frac{1}{\sqrt{\gamma^2(r)-1}} \frac{{\rm d}r}{r^b} \right)^{1/(b-1)} \, ,
\end{equation}
with $\Delta t = t_2 - t_1$, $r_u = (\nu_{\rm m\,2}/\nu_{\rm m\,1})$.
Repeating this step as many times as necessary, we can reconstruct the
entire kinematic evolution of the shock.
The procedure can also be reversed: we can first fit the shock model to the
spectral data, and determine values of $b$ for different time periods.
We then use equation (2) to calculate the respective locations of the
shock at different epochs, and compare these locations with the locations
and speeds inferred from VLBI data.

If we fix the kinematic settings and assume that the shocked feature moves at a speed $\beta_{\rm j}$
along a helical path with amplitude, $A(r)$, frequency, $\omega$, and
parallel wavenumber, $k$, we can reconstruct the time evolution
of the shock location:
\begin{equation}
t(r) = t_0 + \int^r_{r_{0}} \frac{C_2(r)}{k\omega A^2(r) + 
[C_2(r)\beta_{\rm j}^2 - \omega^2 k^2 C_1(r)]^{1/2}}
 dr \,,
\end{equation}
with $C_1(r) = [A^{\prime}_r(r)]^2+1$ and $C_2(r) = C_1(r) + k^2
A^2(r)$.  The form of $A(r)$ may differ, depending on the choice of
the jet geometry. We use $A(r) = A(r_0)r/(a_0 +r)$, corresponding to a
jet with opening half--angle approaching $\arctan[A(r_0)]$, for $r \gg
r_0$. The obtained $t(r)$ can be then checked against $b(t)$ inferred
from the shock model, or used for predicting the light curves
directly.

\section{Examples}

We show here two examples of applying the method outlined above
to radio observations of blazars. 

\subsection{0235+164}

\begin{figure}[t]
\mbox{\psfig{figure=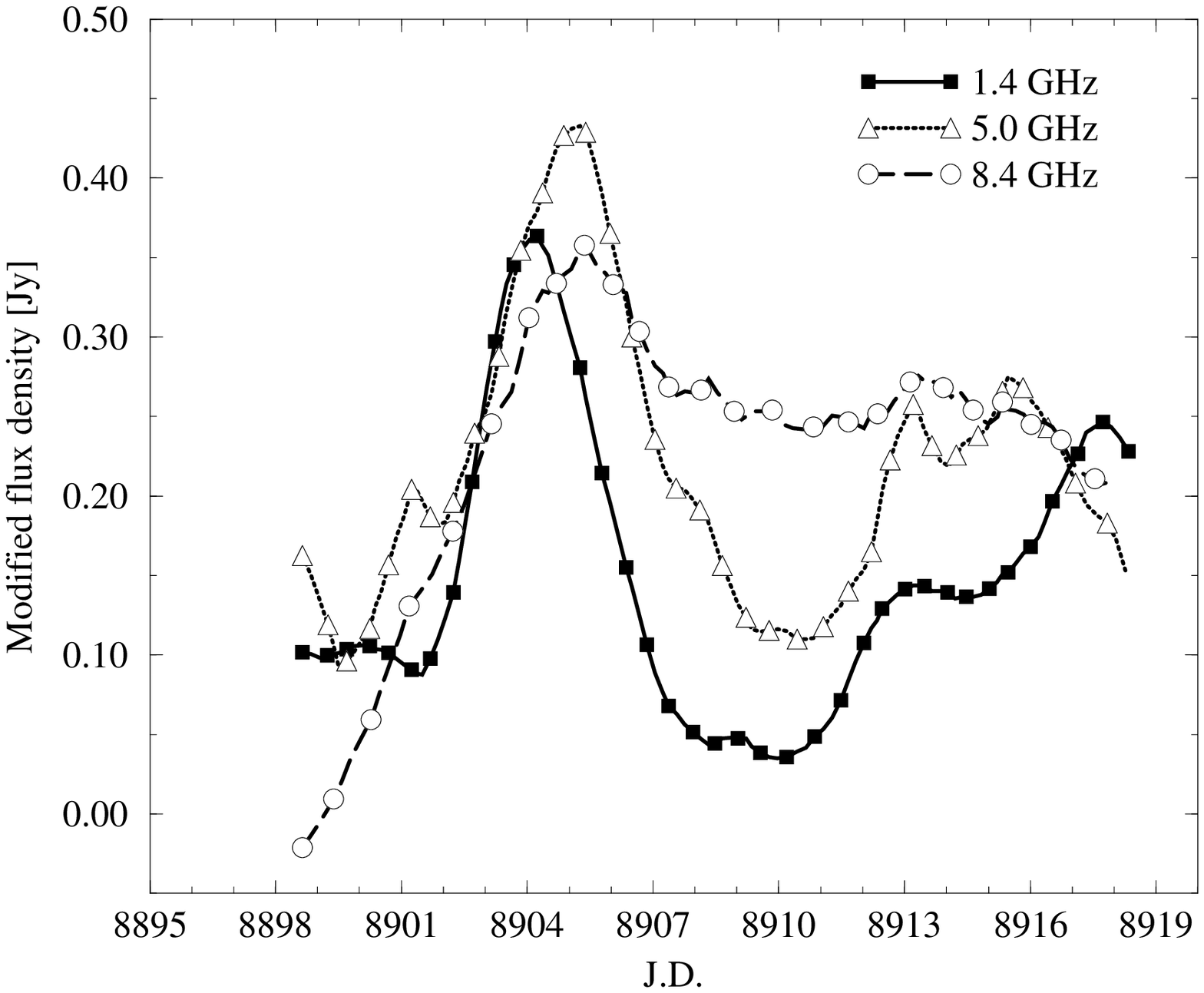,width=0.5\textwidth}
      \psfig{figure=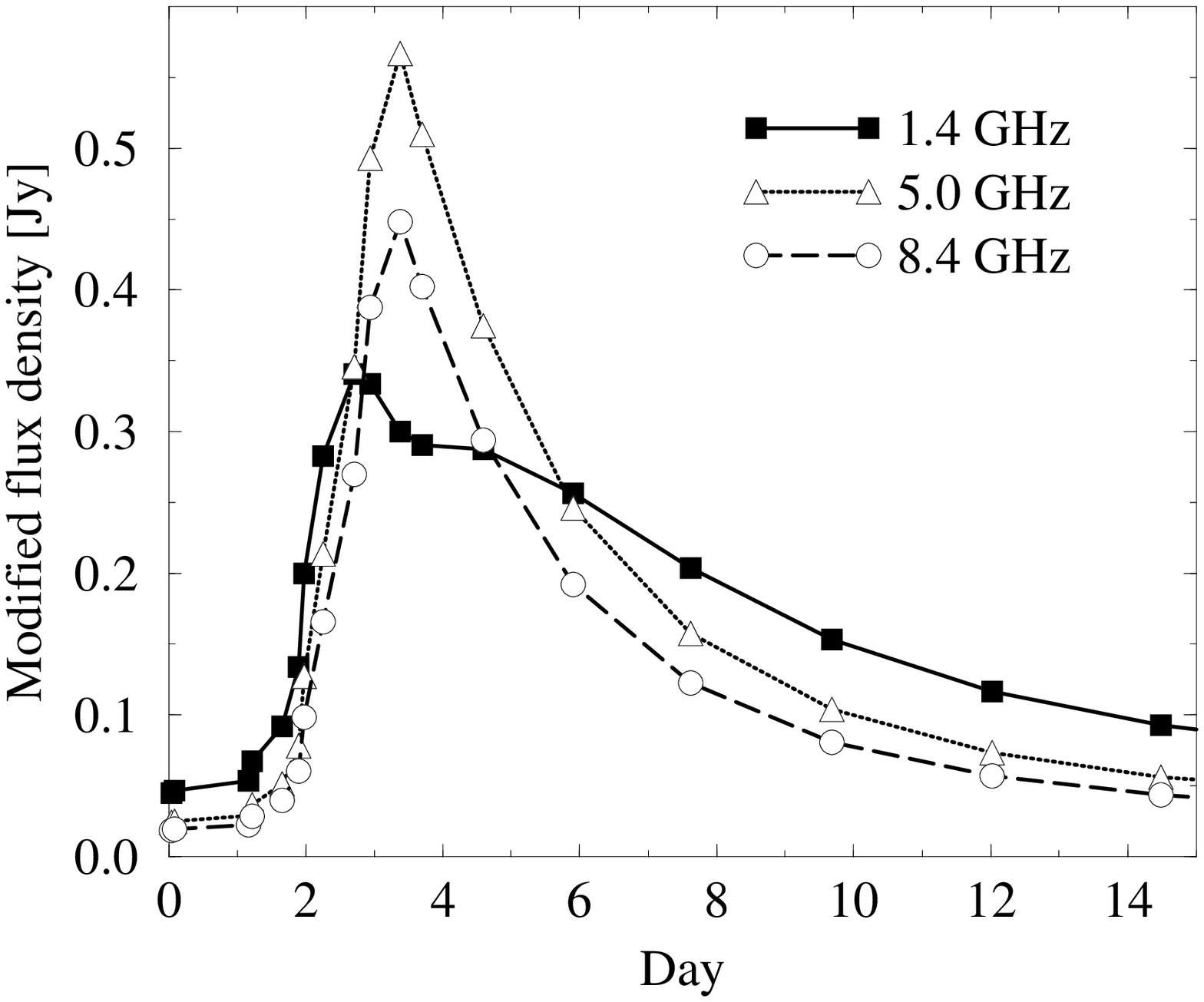,width=0.5\textwidth}}
\caption{Observed (left) and simulated (right) lightcurves of a short--timescale flare in 0235+164. In the observed lightcurves, underlying emission
has been subtracted. The model lightcurves reproduce
the observed timelags, and represent only the first of the
two events (J.D.\,8901--8907) seen in the observed light curves.}
\label{fig-1}
\end{figure}

A short-timescale flare in October 1992 was monitored with the VLA at
1.4, 4.8, and 8.4\,GHz (Kraus, Quirrenbach, Lobanov, et~al., these
proceedings) .  The observed light curves show that the emission first
peaks at 1.4 GHz, and later nearly simultaneously at 4.8 and 8.4\,GHz.
A cross correlation function analysis yields the respective time
lags $\tau^{1.4}_{4.8}= 0.8\pm0.2$\,days, $\tau^{1.4}_{8.4}=
0.7\pm0.2$\,days, and $\tau^{4.8}_{8.4}= -0.2\pm0.2$\,days.  The flare
duration becomes progressively longer at higher frequencies, making
this event rather peculiar. The modified lightcurves obtained after
subtraction of the underlying emission are shown in the left panel of
Figure~\ref{fig-1}.

We discuss elsewhere (Kraus et al. 1999) several possible schemes
capable of explaining the observed peculiarities. One of our proposed
schemes uses a precessing electron--positron beam (see Roland et
al. 1994) with a period $P_0 = 200$\,days and precession angle
$\Omega_0=5.7\deg$.  The kinematics of such a beam is then given by
equation (3), with $A(r_0) = 0.1$\,pc and $a_0 = A(r_0)/\tan\Omega_0 =
1$\,pc. The resulting Doppler factors vary with time, reproducing the
observed lags between the peaks in the lightcurves. For time lags to
be present during a flare, the turnover frequency in the observer's
frame should be within the range of observing frequencies. In
0235+164, we can satisfy this condition by postulating a homogeneous
synchrotron spectrum with spectral index $\alpha=-0.5$ and rest frame
turnover frequency $\nu_{\rm m}^{\prime} = 0.15$\,GHz. Additional
spectral evolution may be required to remove the apparent discrepancy
between the model and observed amplitudes and widths of the flares.

\subsection{3C\,345}

\begin{figure}[t]
\mbox{\psfig{figure=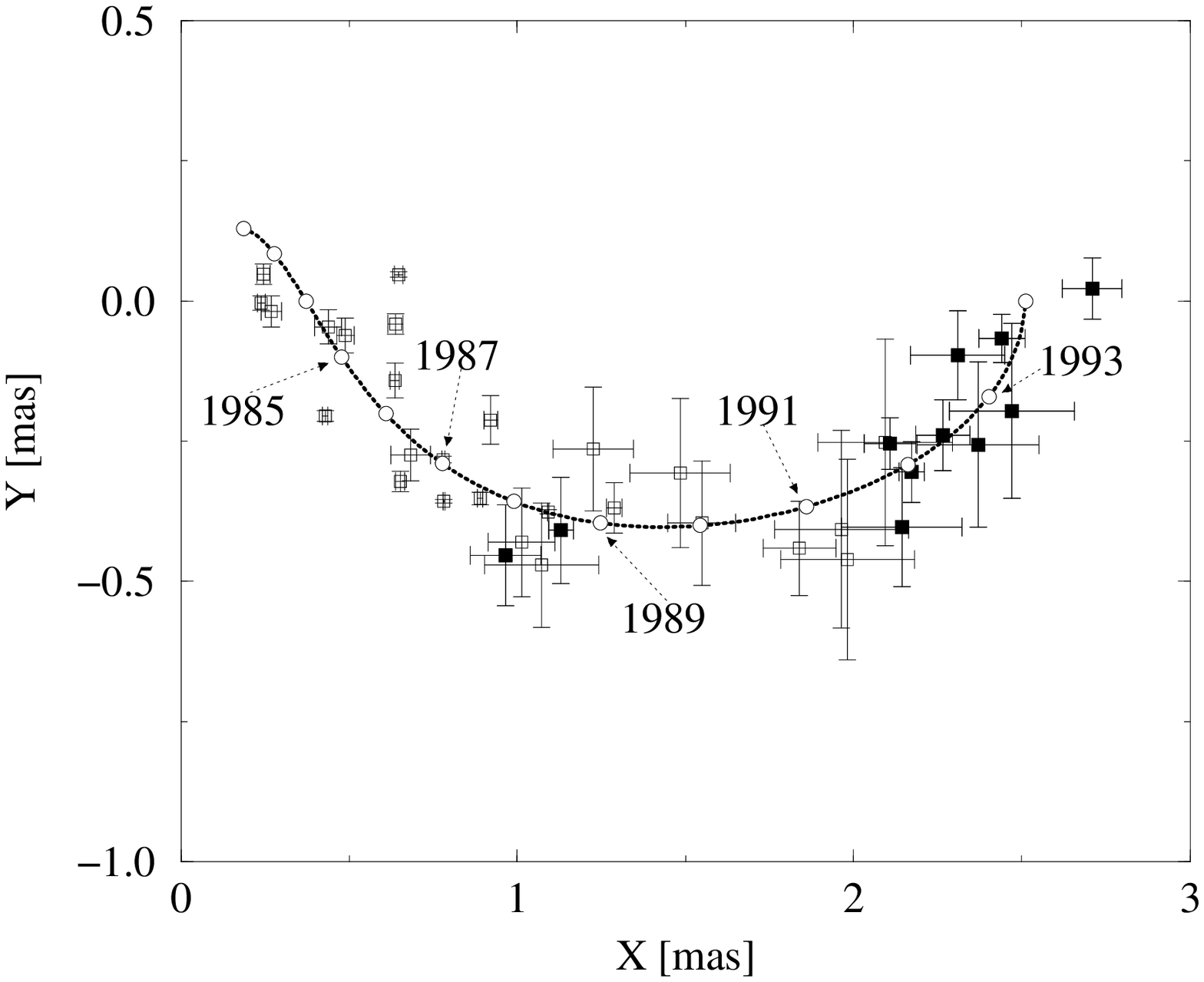,width=0.49\textwidth}
      \psfig{figure=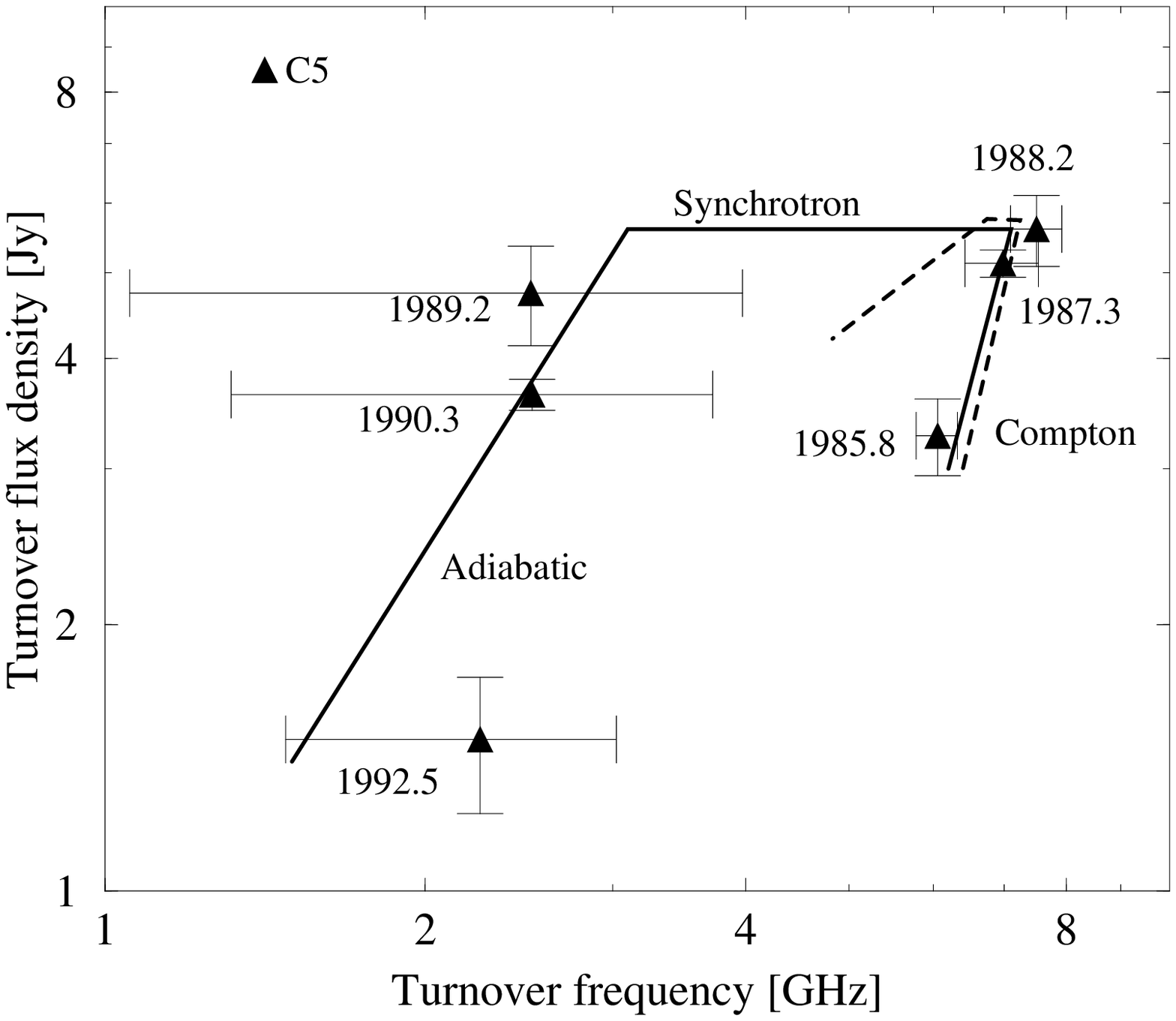,width=0.49\textwidth}}
\caption{Kinematic and spectral evolution of an enhanced emission
region C5 in the jet of 3C\,345. Left panel shows the observed
path of C5 in the plane of the sky; right panel presents the
evolution of the spectral turnover. Solid line in the right
plane shows a fit by the shock model. Dashed line shows how
the fit changes if the spectral evolution of the shock is required
to comply with the observed path of C5.}
\label{fig-2}
\end{figure}

We have studied (Lobanov \& Zensus 1999) spectral changes in the core
and several jet components in 3C\,345, based on the data from a VLBI
monitoring of the source. In the example shown in figure~\ref{fig-2},
we use the observed trajectory (left panel of fig.~\ref{fig-2}) of the
jet component C5 to confront a fit by the shock model to the
variations of $S_{\rm m}$ and $\nu_{\rm m}$ (right panel of
fig.~\ref{fig-2}). In the right panel, the solid line shows a fit by
the shock model, without taking into account the observed kinematics
of C5. When we require the shock model to reproduce $b(t)$ needed to
satisfy the observed path of C5, the fit becomes problematic, at later
stages of the shock evolution. This indicates that, at distances
$>1$\,mas, the shock may have dissipated, and other processes become
main contributors to the emission from C5. Incidentally, the kinematic
and emission properties of other jet features also show evidence for a
change in the emission properties, at distances of 1--1.5\,mas from
the VLBI core of 3C\,345 (Lobanov \& Zensus).

\end{document}